\documentclass[fleqn,twoside]{article}
\usepackage{espcrc2}

\usepackage{graphicx}

\newcommand{\AmS}{{\protect\the\textfont2
  A\kern-.1667em\lower.5ex\hbox{M}\kern-.125emS}}

\title{EPOS Model and Ultra High Energy Cosmic Rays}

\author{T. Pierog\address{Forschungszentrum Karlsruhe, Institut f\"ur 
        Kernphysik, \\ 
        Postfach 3640, Karlsruhe, Germany} 
        and
        K. Werner\address{SUBATECH, Universit\'e de Nantes - IN2P3/CNRS - EMN, \\
        Nantes, France}
       }
\begin{document}

\begin{abstract}
Interpretation of extensive air showers (EAS) experiments results is strongly based on air shower simulations. 
The latter being based on hadronic interaction models, any new model can help for the 
understanding of the nature of cosmic rays. The EPOS model reproducing all major
 results of existing accelerator data (including detailed data of RHIC experiments) 
has been introduced in air shower simulation programs CORSIKA and CONEX few years ago. The new
EPOS~1.99 has recently been updated taking into account the problem seen in EAS development
 using EPOS~1.61. We will show in details the relationship 
between some EPOS hadronic properties and EAS development, as well as the consequences on
the model and finally on cosmic ray analysis.

\vspace{1pc}
\end{abstract}

\maketitle

\section{INTRODUCTION}

Air shower simulations are a very powerful tool to interpret ground
based cosmic ray experiments. However, most simulations are still
based on hadronic interaction models being more than 10 years old.
Much has been learned since, in particular due to new data available
from the SPS and RHIC accelerators. 

In this paper, we discuss air shower simulations based on EPOS, the
latter one being a hadronic interaction model, which does very well
compared to RHIC data~\cite{Bellwied}, and also other
particle physic experiments (especially SPS experiments at CERN). 
But used in air shower simulation
program like CORSIKA~\cite{corsika} or CONEX~\cite{conex}, some results where
in contradiction with KASCADE data~\cite{KASCADE-EPOS}, while it was better for
other experiments~\cite{Glushkov:2007gd}.

Due to the constrains of particle physics, air shower simulations using EPOS 
present a larger number of muons at ground~\cite{Pierog:2006qv}. On the other hand,
we will explain in this paper, how the contrains given by cosmic ray experiments 
can compensate the lack of accelerator data in some given kinematic regions 
(very forward) to improve hadronic 
interaction models and in particular the new EPOS~1.99.

\section{EPOS Model}

One may consider the simple parton model to be the basis of high energy 
hadron-hadron interaction models, which can be seen as an exchange of a 
``parton ladder'' between the two hadrons.
\begin{figure}[th]
\vskip-0.6cm
\begin{center}\includegraphics[width=0.24\textwidth]{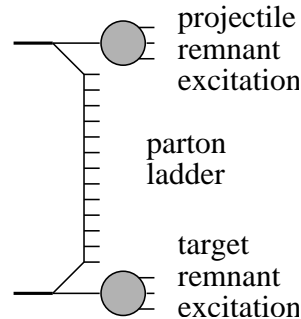}\end{center}
\vskip-1.cm
\caption{Elementary parton-parton scattering: the hard scattering in the
middle is preceded by parton emissions attached to remnants. The remnants 
are an
important source of particle production even at RHIC energies.\label{ladder} }
\vskip-0.65cm
\end{figure}
In EPOS, the term ``parton ladder'' is actually meant to contain two parts \cite{nexus}:
the hard one, as discussed above, and a soft one, which is a purely
phenomenological object, parameterized in Regge pole fashion.

In additions to the parton ladder, there is another source of particle production:
the two off-shell remnants, see fig. \ref{ladder}.
We showed in ref. \cite{nex-bar}
that this {}``three object picture'' can
solve the {}``multi-strange baryon problem'' of conventional
high energy models, see  ref. \cite{sbaryons}.

Hence EPOS is a consistent quantum mechanical multiple scattering approach
based on partons and strings~\cite{nexus}, where cross sections
and the particle production are calculated consistently, taking into
account energy conservation in both cases (unlike other models where
energy conservation is not considered for cross section calculations~\cite{hladik}).
Nuclear effects related to Cronin transverse
momentum broadening, parton saturation, and screening have been introduced
into EPOS~\cite{splitting}. Furthermore, high density effects leading
to collective behavior in heavy ion collisions are also taken into
account~\cite{corona}.

\begin{figure}[htp]
\vskip-0.85cm
\begin{center}
\includegraphics[width=0.43\textwidth]{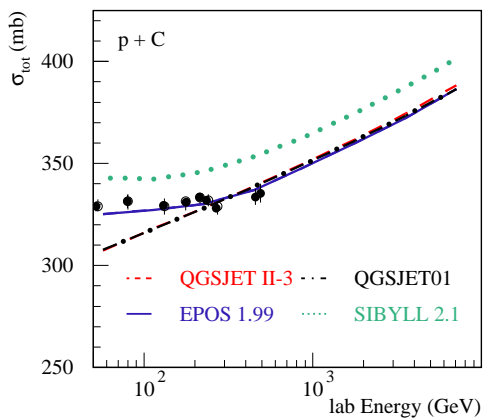}
\end{center}
\vskip-1.35cm
\caption{Total cross section of proton-carbon interactions. EPOS~1.99, QGSJETII, QGSJET01
 and SIBYLL~2.1 hadronic interaction models (lines) are compared to data~\cite{Dersch:1999zg}
 (points) \label{fig-sigC}}
\vskip-.5cm
\end{figure}

Energy momentum sharing and remnant treatment are the key points of the model
concerning air shower simulations because they directly influence the 
multiplicity and the inelasticity of the model. At very high energies or high 
densities, the so-called non-linear effects described in \cite{splitting} are 
particularly important for the extrapolation for EAS and it's one of the parts 
which has been changed in EPOS~1.99.

\subsection{Cross section and inelasticity}

We learned from KASCADE data~\cite{KASCADE-EPOS}, that the energy carried by
hadrons in EPOS~1.61 simulations is too low. It means than the showers are too old
when they reach ground and it was due to a problem in the calculation of the nuclear
cross section and to a too large remnant break-up at high energy (leading to a high 
inelasticity).

To improve the predictive power of the model, the effective treatment of non-linear
effects describe in \cite{splitting} has been made consistent to describe both proton-proton,
hadron-nucleus and nucleus-nucleus data with a unique saturation scale which can be fixed
thanks to proton-proton cross section and Cronin effect in dAu collisions at RHIC. Details
will be published in a dedicated article.

\begin{figure}[htp]
\vskip-0.85cm
\begin{center}
\includegraphics[width=0.43\textwidth]{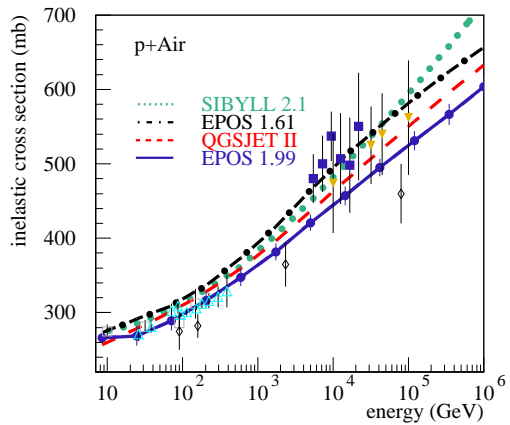}
\end{center}
\vskip-1.35cm
\caption{Inelastic cross section of proton-air interactions. EPOS~1.99, QGSJETII, EPOS~1.61
 and SIBYLL~2.1 hadronic interaction models (lines) are compared to data of air shower experiment (points). \label{fig-sigAir}}
\vskip-.5cm
\end{figure}

The EPOS~1.99 (full line) proton-carbon total cross section is shown Fig~\ref{fig-sigC}.
It is now in very good agreement with the data~\cite{Dersch:1999zg} and with 
the other hadronic interaction model used
for air shower physics QGSJET01~\cite{qgsjet01} (dash-dotted line), QGSJETII~\cite{qgsjetII} (dashed line) 
and SIBYLL~\cite{sibyll} (dotted line). In fig~\ref{fig-sigAir}, 
the extrapolation to proton-air data up to the highest energies is shown in comparison with 
measurement from cosmic ray experiments. The error bar for EPOS~1.99 represents the uncertainty 
due to the definition of the inelastic cross section as measured by cosmic ray 
experiments\footnote{The difference between the top and the bottom of the error bars is the part 
of the cross-section where secondary particles are produced without changing the projectile 
(target diffraction). Cross section of other models includes this target diffraction (top of error bars).}. In comparison with EPOS~1.61 (dash-dotted line), the EPOS~1.99 cross section
 has been notably reduced.

\subsection{Particle production}

\begin{figure}[htbp]
\vskip-1.cm
{\par \hfill
\begin{center}
\includegraphics[  scale=0.35]{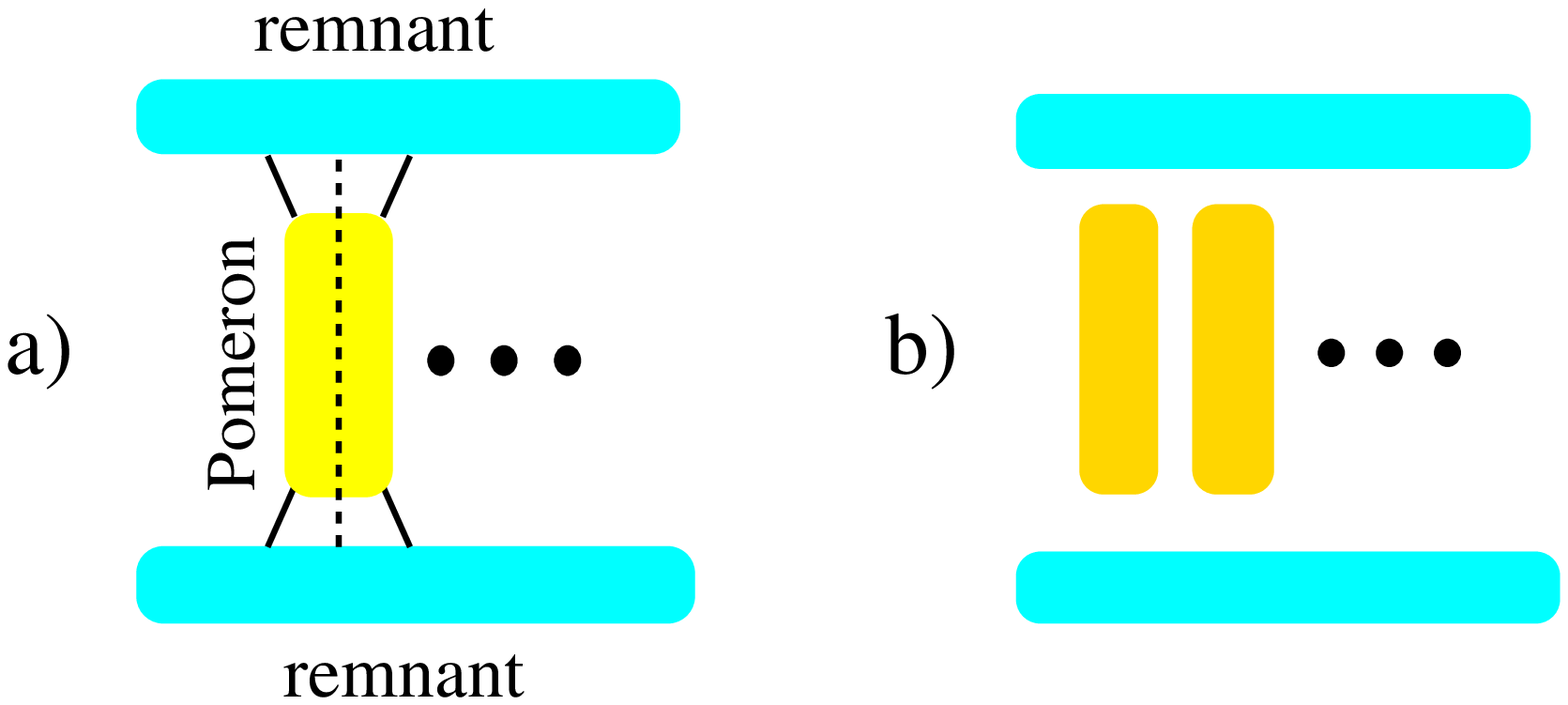}\hspace{1cm}
\includegraphics[  scale=0.45]{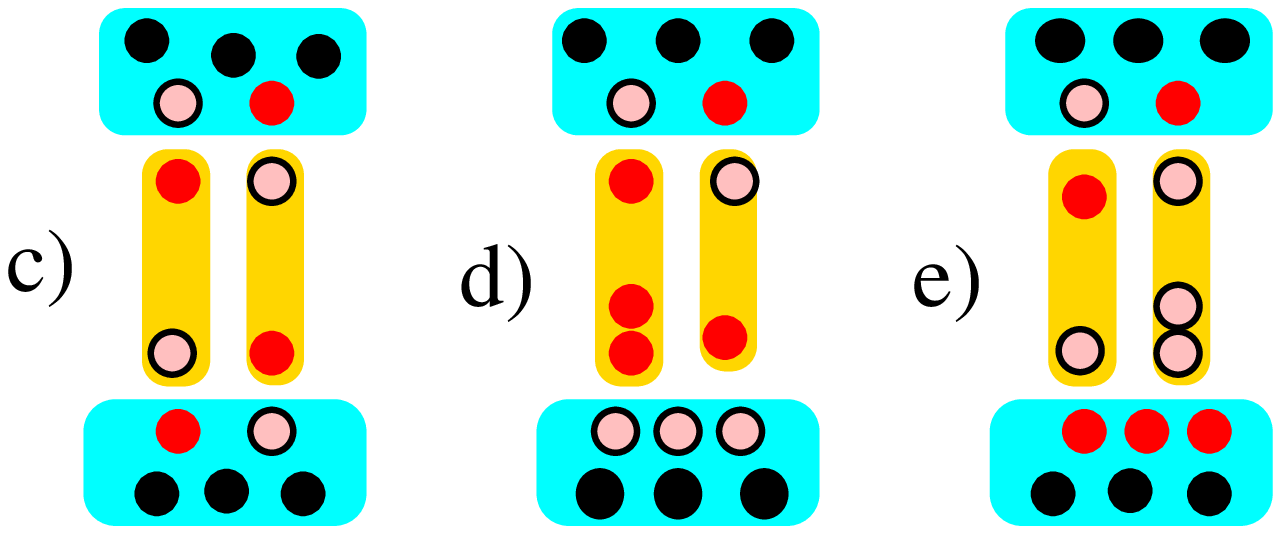}
\end{center}\hfill
\par}
\vskip-1.5cm
\caption{ a) Each cut Pomeron is regarded as two strings b). 
c) The most simple and frequent collision
configuration has two remnants and only one cut Pomeron represented
by two $\mathrm{q}-\overline{\mathrm{q}}$ strings. d) One of the
$\overline{\mathrm{q}}$ string ends can be replaced by a $\mathrm{qq}$
string end. e) With the same probability, one of the $\mathrm{q}$
string ends can be replaced by a $\overline{\mathrm{q}}\overline{\mathrm{q}}$
string end. 
\label{nexus1}}
\end{figure}

Thanks to a Monte Carlo, first the collision configuration is determined:
i.e. the number of each type of Pomerons exchanged between the projectile
and target is fixed and the initial energy is shared between the Pomerons
and the two remnants. Then particle production is accounted from two
kinds of sources, remnant decay and cut Pomeron. A Pomeron may be
regarded as a two-layer (soft) parton ladder attached to projectile
and target remnants through its two legs. Each leg is a color singlet,
of type q$\overline{\mathrm{q}}$ , qqq or 
$\overline{\mathrm{q}}$$\overline{\mathrm{q}}$$\overline{\mathrm{q}}$ from the
sea, 
and then each cut Pomeron is regarded as two strings, cf. Fig.~\ref{nexus1}a. %

It is a natural idea to take quarks and antiquarks from the sea as
string ends for soft Pomeron in EPOS, because
an arbitary number of Pomerons may be involved. 
In addition to this soft Pomerons,
hard and semihard Pomerons are treated differently. 

\begin{figure}
\begin{center}\includegraphics[width=0.43\textwidth]{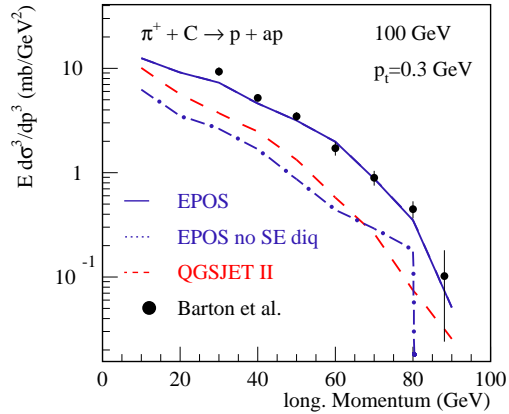}\end{center}
\vskip-1.35cm
\caption{Model comparison: longitudinal momentum distributions of pion carbon 
collisions at 100 GeV from EPOS with (full) or without (dash-dotted)
sting-end diquarks and QGSJETII (dashed) compared to data~\cite{barton}.}
\label{fig-se}
\vskip-0.75cm
\end{figure}

Thus, besides the three valence quarks,
each remnant has additionally quarks and antiquarks to compensate
the flavours of the string ends, as shown in fig.~\ref{nexus1}c. 
According to its number of quarks and antiquarks, to the phase space, and to
an excitation probability, a remnant decays into mesons, (anti)baryons \cite{nex-bar}. 
Furthermore, this process leads
to a baryon stopping phenomenon in which the baryon number 
can be transfered from the remnant to the string ends (for instance 
in~\ref{nexus1}d, depending
on the process, the $3\overline{\mathrm{q}}+3\mathrm{q}$ can be
seen as 3 mesons or a baryon-antibaryon pair). 

In case of meson projectile, this
kind of diquark pair production at the string ends leads to an increase of the
(anti)baryon production in the forward production in agreement with low energy
pion-nucleus data~\cite{barton} as shown fig.~\ref{fig-se}. As a consequence it 
is part of the larger number of muons in EAS simulations with EPOS.

Compared to EPOS~1.61, EPOS~1.99 has a reduced excitation probability at high 
energy, increasing the number of protons in the forward direction and reducing 
the inelasticity.

\section{AIR SHOWERS}

In the following, we discuss air shower simulations, based on the shower
programs CONEX, using the old
EPOS~1.61 (dash-dotted line), the new EPOS~1.99 (full line) and QGSJETII(dashed line) (as a reference) as high 
energy hadronic interaction model in KASCADE experiment energy range.

\begin{figure}[htp]
\vskip-0.7cm
\begin{center}\includegraphics[width=0.47\textwidth]{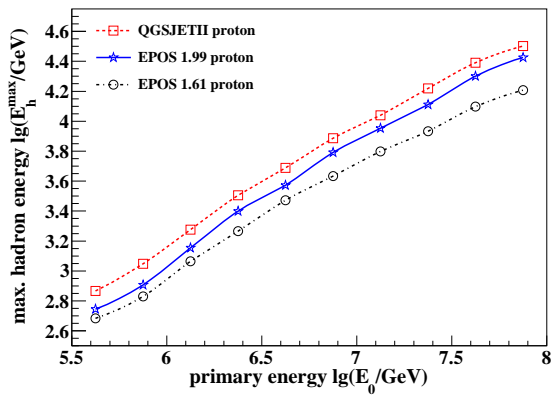} \end{center}
\vskip-1.15cm
\caption{Maximum hadron energy as a function of the primary energy for proton 
induced showers using EPOS~1.99 (full line), EPOS~1.61 (dash-dotted line) and 
QGSJETII (dashed line) as high energy hadronic interaction model.\label{fig-Emax}}
\vskip-0.7cm
\end{figure}

The effect of the reduced cross section and inelasticity is clearly visible on
the maximum energy of hadrons at ground as shown fig.~\ref{fig-Emax}. The shower
being younger at ground with EPOS~1.99, the maximum energy is up to 60\% higher than 
in the previous release 1.61. The results are now close to QGSJETII results but with
 a different slope due to a different elongation rate. As a consequence, EPOS~1.99
does not have the problems pointed out in \cite{KASCADE-EPOS} anymore and should be
compatible with KASCADE data.

\section{Summary}

EPOS is a new interaction model constructed on a solid theoretical
basis. It has been tested very carefully against all existing hadronic
data, also those usually not considered important
for cosmic rays. In EAS simulations, EPOS provides more muons
than other models, which was found to be linked to an increased diquark
production in both string ends and string fragmentation. The new EPOS~1.99 has 
a reduced cross section and inelasticity compared to the previous EPOS~1.61 which
leads to deeper shower development. This would solve
the problem with KASCADE data. But since the number of muons and the elongation rate 
are different than in the other models, the resulting analysis will be significantly
 different.


\begin{thebibliography}{9}
\bibitem{Bellwied}
R.~Bellwied. {\em Acta Phys. Hung.}, A27:201--204, 2006.

\bibitem{corsika}
D.~Heck et~al. FZKA-6019, 1998.

\bibitem{conex}
T.~Bergmann et~al. {\em Astropart. Phys.}, 26:420--432, 2007.

\bibitem{KASCADE-EPOS}
W.~D.~Apel et~al., {\em J. Phys. G: Nucl. Part. Phys.} 36:035201, 2009

\bibitem{Glushkov:2007gd}
  A.~V.~Glushkov et~al., 
  {\em JETP Lett.}, 87:190 2008.

\bibitem{Pierog:2006qv}
T.~Pierog and K.~Werner. {\em Phys. Rev. Lett.} 101:171101 2008.

\bibitem{nexus}
H.~J. Drescher et~al. {\em Phys. Rept.}, 350:93--289, 2001.

\bibitem{nex-bar}
F.~M. Liu et~al. {\em Phys. Rev.}, D67:034011, 2003.

\bibitem{sbaryons}
M.~Bleicher et~al. {\em Phys. Rev. Lett.}, 88:202501, 2002.

\bibitem{hladik}
M.~Hladik et~al. {\em Phys. Rev. Lett.}, 86:3506--3509, 2001.

\bibitem{splitting}
K.~Werner et~al. {\em Phys. Rev.}, C74:044902, 2006.

\bibitem{corona}
K.~Werner. {\em Phys. Rev. Lett.}, 98:152301, 2007.

\bibitem{Dersch:1999zg}
  U.~Dersch et al. 
 {\em  Nucl. Phys. }  B579:277, 2000.

\bibitem{qgsjet01}
N.~N. Kalmykov, S.~S. Ostapchenko, and A.~I. Pavlov,
Nucl. Phys. Proc. Suppl. 52B (1997) 17--28.

\bibitem{qgsjetII}
S.~Ostapchenko. {\em Phys. Rev.}, D74:014026, 2006.

\bibitem{sibyll}
R.~Engel, T.~K. Gaisser, P.~Lipari, and T.~Stanev,
in Proceedings of 26th ICRC (Salt Lake City)
  vol.~1, p.~415,
1999.


\bibitem{barton}
D.~S. Barton et~al. {\em Phys. Rev.}, D27:2580, 1983.

\end{thebibliography}
\end{document}